\documentclass[prx,letterpaper,nobalancelastpage,twocolumn,superscriptaddress,nofootinbib]{revtex4}

\usepackage{graphicx}
\usepackage{amsmath}
\usepackage{amssymb}
\usepackage[english]{babel}

\usepackage{xcolor}
\usepackage[version=4]{mhchem}
\usepackage[hidelinks]{hyperref}
\usepackage{dsfont}
\usepackage{multirow}
\usepackage{makecell}

\usepackage{color}

\usepackage{soul}
\newcommand{\ket}[1]{{| #1 \rangle}}

\newcommand{\UIUC}{Department of Physics, The University of Illinois at Urbana-Champaign, Urbana, IL 61801, USA}
\newcommand{\UC}{Pritzker School of Molecular Engineering, University of Chicago, Chicago, IL 60637, USA}
\newcommand{\LMUa}{Faculty of Physics, Ludwig-Maximilians-University of Munich, Munich, Germany}
\newcommand{\LMUb}{Munich Center for Quantum Science and Technology, Munich, Germany}
\newcommand{\MPQ}{Max-Planck Institute for Quantum Optics, Garching, Germany}

\renewcommand{\cite}[1]{\mbox{\citep{#1}}}

\addto\captionsenglish{}
\addto\captionsenglish{}

\begin{document}

\title{Quantum networks with neutral atom processing nodes}
\author{Jacob P. Covey}\email{jcovey@illinois.edu}
\affiliation{\UIUC}
\author{Harald Weinfurter}\email{h.w@lmu.edu}
\affiliation{\LMUa}
\affiliation{\LMUb}
\affiliation{\MPQ}
\author{Hannes Bernien}\email{bernien@uchicago.edu}
\affiliation{\UC}

\begin{abstract}
Quantum networks providing shared entanglement over a mesh of quantum nodes will revolutionize the field of quantum information science by offering novel applications in quantum computation, enhanced precision in networks of sensors and clocks, and efficient quantum communication over large distances. Recent experimental progress with individual neutral atoms demonstrates a high potential for implementing the crucial components of such networks. We highlight latest developments and near-term prospects on how arrays of individually controlled neutral atoms are suited for both efficient remote entanglement generation and large-scale quantum information processing, thereby providing the necessary features for sharing high-fidelity and error-corrected multi-qubit entangled states between the nodes. We describe both the functionality requirements and several examples for advanced, large-scale quantum networks composed of neutral atom processing nodes.

\end{abstract}
\maketitle

\section{Introduction and grand vision}\label{vision}
\begin{figure*}[t!]
    \centering
    \includegraphics[width=16cm]{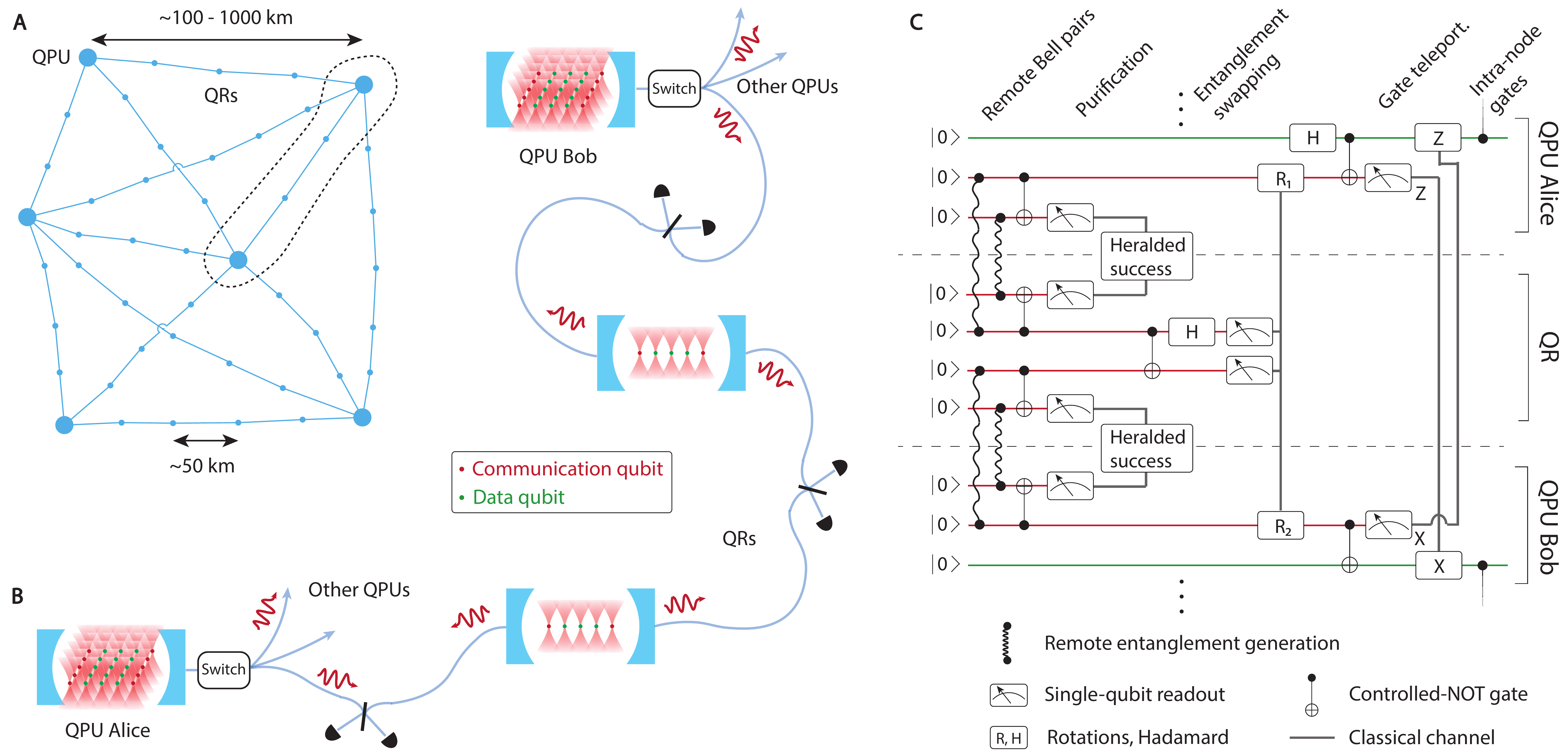}
    \caption{
        \textbf{Vision for a repeater-enabled long-distance network between neutral atom quantum processing units (QPUs)}. \textbf{A} A possible network architecture with several large-scale QPU nodes connected by links containing intermediate Quantum Repeater (QR) nodes. \textbf{B} Atom arrays in optical cavities interfaced with optical photons serve as the nodes; QRs are similar, possibly smaller versions of the end-node QPUs. Repeaters can be spaced by approximately 50 km and contain multiple atoms for multiplexing, entanglement swapping, and other deterministic logic operations such as entanglement purification. Red atoms denote communication qubits and green atoms denote data qubits. \textbf{C} An example circuit to generate a purified Bell pair between two QPUs via a single QR. Successful entanglement purification is heralded, but failure requires that the link be reestablished. Gate teleportation is needed to transfer the Bell pair from the communication qubits to the data qubits. The operations are shown in the legend. Readout is in the z-basis unless stated otherwise. Note that we have assumed that communication qubits can also serve as data qubits in the QRs but not in the QPUs. 
    }
    \label{Fig1}
\end{figure*}

The development of large-scale quantum networks~\cite{Cirac1997,Wehner2018,Kimble2008} will usher in an era of novel applications of quantum technology, which include cryptographically-secured communication~\cite{Pirandola2019}, distributed or blind quantum computing~\cite{Jiang2007}, and sensor and clock networks with precision approaching the fundamental quantum limit~\cite{Komar2014,Gottesman2012}. Such a network will consist of a mesh of quantum nodes, which we refer to as ``quantum processing units" (QPUs), interconnected with quantum links capable of efficient distribution of quantum states over the whole system (see Fig.~\ref{Fig1}A). The quantum network will in many ways operate analogously to the classical internet in which classical computers or sensors constitute each node, but it will also face unique challenges due to the fragility of quantum information and the inability to clone a quantum state for signal amplification~\cite{Wootters1982}.

In spite of significant progress in recent years, the realization of a large-scale network poses a number of challenges. First, the quantum systems must provide an optical interface that connects their states with quantum states of light to enable remote entanglement generation (REG) over a link (see Fig.~\ref{Fig1}B). If the distance between nodes exceeds a threshold, a quantum repeater scheme must be employed~\cite{Briegel1998}, in which entanglement is distributed between distant nodes by first sharing entanglement over intermediate links with quantum repeater (QR) stations that are then connected together (see Fig.~\ref{Fig1}A). Second, as the REG process to share entanglement over a link is stochastic, it does not always succeed and needs to be repeated until successful, which can take a significant amount of time. Therefore, quantum memories based on long-lived states are required to maintain the quantum states at the nodes and QR stations with high fidelity. Third, for connecting the links using entanglement swapping~\cite{Zukowski1993,Pompili2021}, deterministic quantum logic operations are required at QRs and at the nodes (see Fig.~\ref{Fig1}C). Fourth, since the remote entanglement has limited fidelity that is even further reduced when connecting many intermediate links, ``entanglement purification" across the entire link is required~\cite{Bennett1996a,Dur2003,Kalb2017}. Purification is also a stochastic process; if it fails, the whole process on this part of the link has to be repeated (see Fig.~\ref{Fig1}C), thus requiring even longer storage times in the quantum memories -- often approaching the second-scale. Eventually, active error correction~\cite{Knill1998,Knill2005} will be required to enable the requisite long coherence times and to store the distributed states. 

Individual neutral atoms have the potential to implement many highly desirable features of quantum network nodes including efficient light-matter interfaces -- potentially at telecom wavelengths~\cite{Uphoff2016,Covey2019b, Menon2020} -- based on optical cavities~\cite{Kimble2008,Reiserer2015}, minute-scale coherence and memory times~\cite{Young2020,Barnes2022,Ma2022,Jenkins2022}, multi-qubit processing capabilities~\cite{Saffman2010,Levine2019,Graham2019,Madjarov2020}, scalability to hundreds of qubits~\cite{Ebadi2021}, and even high-fidelity mid-circuit readout~\cite{Deist2022b,Singh2022,Graham2023}. Accordingly, we envision long-distance networks with QPUs and QRs as arrays of individually-controlled atoms within optical cavities (see Fig.~\ref{Fig1}B). In general, the QPUs and QRs could contain two types of qubits: communication qubits and data qubits, which are used to create remote Bell pairs and to process quantum information within the node, respectively. Here, we present a Perspective on the combination of recent advances on research with individual neutral atoms, from which near-term developments will constitute a large step towards realizing our vision for robust quantum networks.

\begin{figure*}[t!]
    \centering
    \includegraphics[width=14cm]{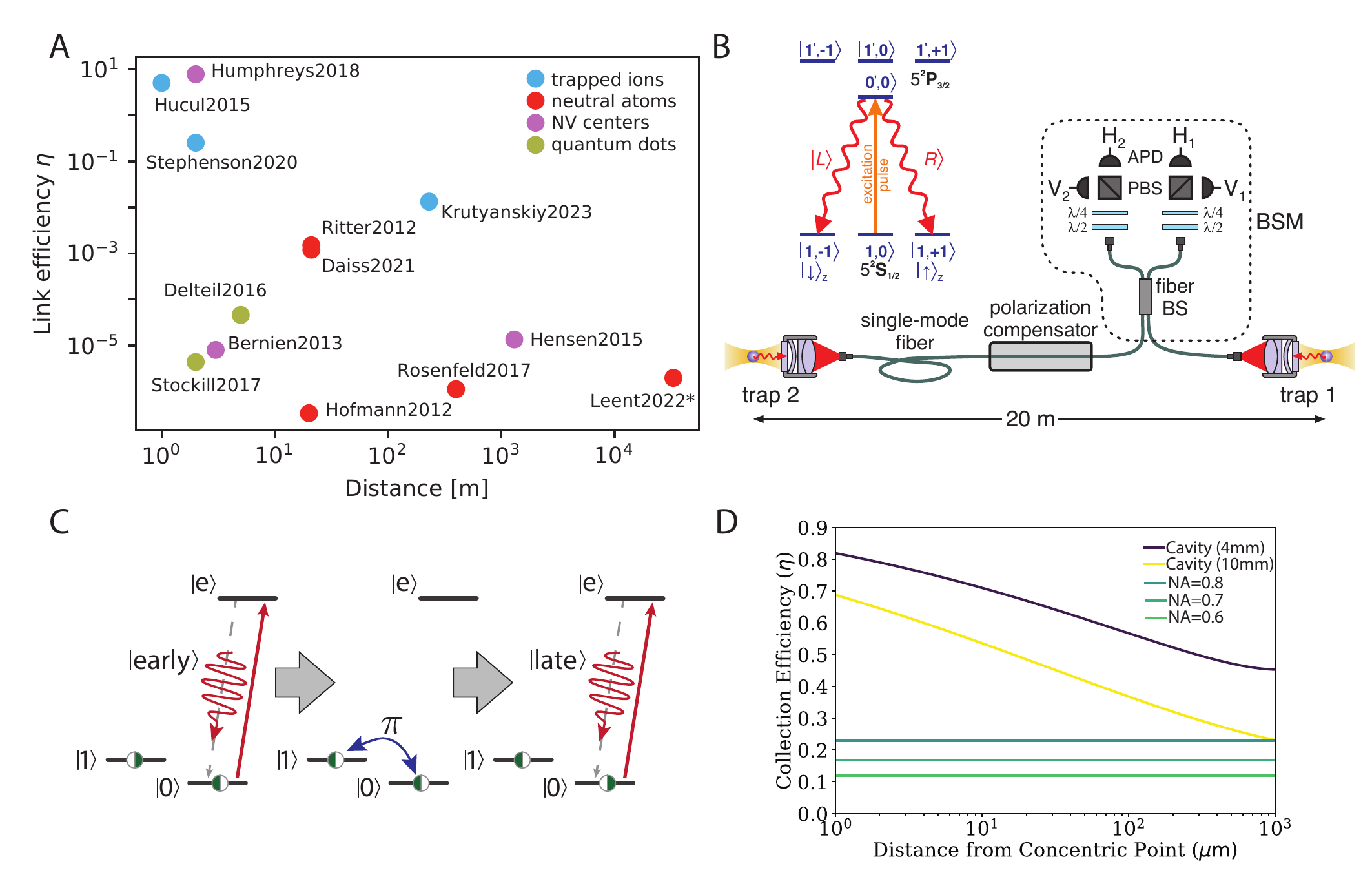}
    \caption{\textbf{Long-distance remote entanglement generation.}
        \textbf{A} Overview of experimental REG demonstrations between two nodes. Shown is the link efficiency $\eta$ in dependence of the separation between the nodes. ${}^*$For Leent2022 the direct distance was $400\,$m with an additional $32\,$km of coiled optical fiber. \textbf{B} Sketch of the setup used in~\cite{Hofmann2012}. At each node the atomic qubit is entangled with a photon in the polarization basis (see level diagram). The photons are collected into a single mode fiber and interfered on a beamsplitter (fiber BS). A joint detection of the photons swaps the entanglement to the two atoms and heralds success of the protocol. Adapted from~\cite{Hofmann2012}. \textbf{C} The atomic qubit can also be entangled with a photon in the time-bin basis. The protocol consists of the preparation of a qubit superposition, followed by two qubit-state-selective excitations separated by a qubit $\pi$-pulse. \textbf{D} Theoretical collection efficiency $\eta$ of photons emitted by a single atom using microscope objectives with various numerical apertures (NA) or near-concentric cavities with different lengths. Adapted from~\cite{Young2022b}.
    }
    \label{Fig2}
\end{figure*}

Although we focus only on neutral atoms, we note that many hardware platforms are actively being pursued for the realization of this vision. Nitrogen-vacancy centers in diamond is perhaps the most advanced platform (see e.g. Refs.~\cite{Awschalom2018,Bernien2013,Hensen2015,Humphreys2018,Pompili2021}), and other defects such as silicon- and germanium-vacancy centers in diamond offer attractive features~\cite{Sipahigil2016,Bradac2019,Bhaskar2020,Raha2020}. Other solid-state systems such as semiconductor quantum dots~\cite{Elzerman2004,Hanson2007,Hanson2008} and rare earth ion-doped crystals~\cite{Zhong2017,Dibos2018,Zhong2019,Chen2020,Kindem2020,Craiciu2021,Ulanowski2022,Uysal2023} are showing great promise. Trapped ions have much in common with neutral atoms, and are one of the most mature platforms for quantum science more generally (see e.g. Refs.~\cite{Moehring2007,Olmschenk2009,Hucul2015,Stephenson2020,Nichol2022,Krutyanskiy2023}). The progress towards a quantum network is illustrated in Fig.~\ref{Fig2}A across all platforms, which shows the current record ``link efficiency" -- the ratio of entanglement generation rate to decoherence rate (see below and Refs.~\cite{Monroe2014,Hucul2015}) -- versus link distance.

\section{Remote entanglement generation}\label{REG} 
The generation of remote entanglement between distant matter qubits is best mediated by light. For example, two distant atomic nodes can be entangled by first creating an atom-photon entangled state at one node and then storing the  photon in the other~\cite{Ritter2012}. Note that in this generic scheme the entangled state fidelity is vulnerable to photon loss and it is challenging to implement a heralding signal indicating the successful generation of entanglement. 

Alternatively, for atom-cavity coupled nodes (see also section~\ref{REG:cavity}) a photon that is sequentially reflected off two distant nodes can be used to perform a non-local Bell measurement~\cite{Welte2021} or mediate a non-local gate between the atoms~\cite{Daiss2021} which can both create long-distance entanglement between the nodes including a heralding signal. 

Finally, if each of the neighboring nodes emits a photon that is entangled with a matter qubit at the node, one is able to swap the matter-light entanglement to the matter qubits using a joint measurement in the Bell basis (Bell-state measurement, BSM) on the photons~\cite{Zukowski1993,Cabrillo1999,Barrett2005}. 
The detection of the photons at the BSM also provides the heradling signal. In this perspective we focus on this entanglement swapping protocol for REG but note that the techniques described here are also applicable to other REG protocols.

\subsection{REG based on entanglement swapping}
The protocol starts by preparing the atom-photon entangled state at each node $\ket{\psi}=1/\sqrt{2}(\ket{0}\ket{\alpha}+\ket{1}\ket{\beta})$, where \{$\ket{0}$, $\ket{1}$\} are the states of the atomic qubit and \{$\ket{\alpha}$, $\ket{\beta}$\} are two orthogonal states of the emitted photon, for instance two different polarization states or different emission time bins (see Fig.~\ref{Fig2}B and C). The photons are sent to a middle node and overlapped on a beamsplitter. 
For indistinguishable photons, the anti-symmetric state $\ket{\Psi^-}=1/\sqrt{2}(\ket{\alpha_1}\ket{\beta_2}-\ket{\beta_1}\ket{\alpha_2})$ results in detection of one photon in each of the two output ports of the beamsplitter which projects the common state of the two atoms into an entangled state $\ket{\Psi^-}=1/\sqrt{2}(\ket{0}\ket{1}-\ket{1}\ket{0})$. At the same time the photon detection acts as a heralding signal. It indicates the success of the REG and allows the user to disregard unsuccessful attempts. This makes the protocol robust against photon losses which then in principle only affect the success rate of the REG but not the fidelity of the entangled state. Besides the anti-symmetric photonic state in the beamsplitter output ports, there are three symmetric states in which both photons exit into the same output port. Analysis of the respective degree of freedom (e.g. polarization) unveils one of them~\cite{Hofmann2012} which projects the nodes the nodes in the symmetric state $\ket{\Psi^+}=1/\sqrt{2}(\ket{0}\ket{1}+\ket{1}\ket{0})$. The other two photonic Bell-states give the same detection signal and cannot be further distinguished~\cite{Calsamiglia2000}. Hence, the maximally achievable success probability of the BSM is 50\%.  In Fig.~\ref{Fig2}~B the additional polarization beam splitters (PBS) in the output port of the fiber beamsplitter, enable such detection of the antisymmetric and symmetric Bell states.
 
\subsection{REG experimental demonstrations}

The entanglement swapping protocol has been realized with several different physical platforms (see Fig.~\ref{Fig2}~A) including trapped ions~\cite{Moehring2007,Hucul2015,Stephenson2020,Krutyanskiy2023}, NV centers in diamond~\cite{Bernien2013,Hensen2015,Humphreys2018}, quantum dots~\cite{Stockill2017,Delteil2016} and neutral atoms~\cite{Hofmann2012,Rosenfeld2017,Leent2022}. One way to compare the performance of these experiments is  defining the link efficiency $\eta=\gamma_\text{ent}/\gamma_\text{dec}$ as a figure of merit, which relates the entanglement rate $\gamma_\text{ent}$ to the decoherence rate $\gamma_\text{dec}$ of the entangled state~\cite{Hucul2015,Humphreys2018} (see Subsection \ref{sec:link-eff}). Figure~\ref{Fig2}~A gives an overview of 2-node quantum network experiments, showing $\eta$ in dependence of the separation between the network nodes. For short distances of a few meters, $\eta>1$ has been achieved~\cite{Hucul2015, Humphreys2018}, i.e., entanglement could be generated faster than its decoherence time. As the distance between the nodes becomes longer, $\eta$ decreases due to optical losses reducing the entanglement rate. 

For REG experiments a variety of photonic state encodings have been used, ranging from polarization~\cite{Moehring2007,Hofmann2012} (Fig.~\ref{Fig2}~B), energy~\cite{Olmschenk2009}, time-bin~\cite{Bernien2017,Hensen2015}~(Fig.~\ref{Fig2}~C), and number-state encoding~\cite{Stockill2017,Humphreys2018}. While the measurement-based protocol in principle works for all these different degrees of freedom, certain encodings can be more preferable depending on the specific physical implementation of the network. For instance, time-bin encodings have proven to be robust to polarization drifts in optical fibers. For number state encoding, the atomic state is entangled with the Fock states of the photon mode $\ket{0}$ and $\ket{1}$. Only a single photon coming from either node needs to be detected at the BSM station to swap the entanglement to the atoms. For large node separations this results in faster entanglement rates but needs interferometric stability of the optical channels~\cite{Stockill2017,Humphreys2018}.

\subsection{Light-matter interfaces for neutral atoms}
\label{REG:cavity}

\begin{figure}[t!]
    \centering
    \includegraphics[width=8.6cm]{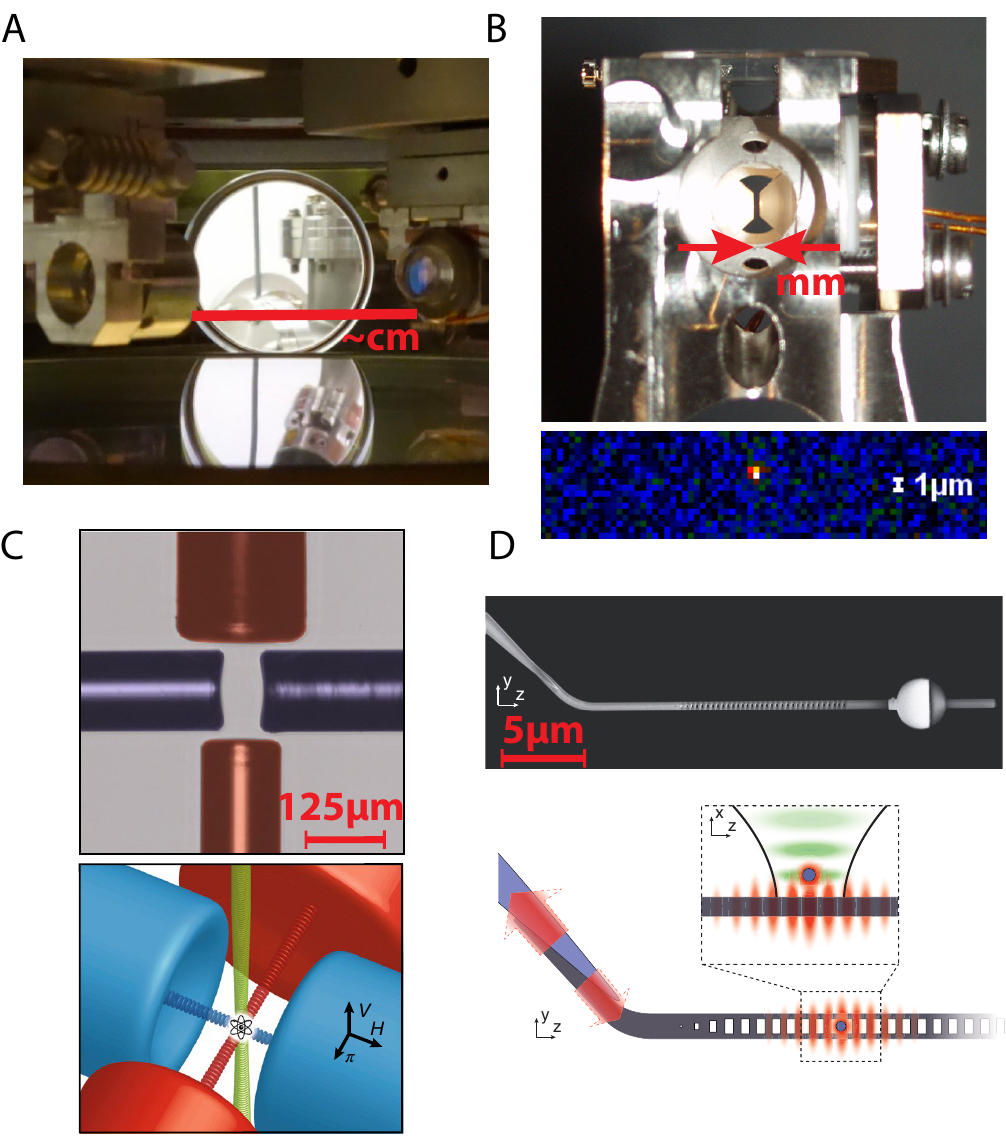}
    \caption{
        \textbf{Cavity architectures for enhancing REG rates.} \textbf{A} Centimeter-scale near-concentric cavity. Adapted from~\cite{Davis2020}. \textbf{B} A millimeter-scale Fabry-P\'erot cavity with the fluorescence of a single atom as inset. Adapted from~\cite{Hacker2019}. \textbf{C}  A crossed-fiber cavity in which the fiber facets are coated and shaped to provide high reflectivity mirrors. Adapted from~\cite{Brekenfeld2020}. \textbf{D} Photonic crystal cavities confine light to sub-$\lambda^3$ volumes, where $\lambda$ is the wavelength of the light. Placing an atom into this cavity mode leads to coupling with $C>1$. Adapted from~\cite{Tiecke2014}.
    }
    \label{fig:cavities}
\end{figure}

Individual neutral atoms have been used for quantum network demonstrations in a variety of different architectures. The main distinguishing feature of these experiments is the method by which the light-matter interface is realized. For instance, photons that are emitted by the atoms into free space can be collected using high-numerical aperture optics. The collection efficiency in this approach is limited by geometric constraints and is typically on the order of a few percent even when using microscope objectives with high numerical apertures (typical NA$\sim 0.5$). This free-space approach was used for example with polarization-entangled photons that were collected from atoms in optical dipole traps~\cite{Hofmann2012} (see Fig.~\ref{Fig2} B). With a separation of 20~m between the atoms, the entanglement rate for this experiment was $9\,$mHz, primarily limited by the collection efficiency of the photons. This collection efficiency can be significantly increased by coupling atoms to optical cavities for which emission into a specific mode can be enhanced (see Fig.~\ref{Fig2}~D). For instance, an entanglement rate of $30\,$Hz over a similar distance of $21\,$m has been achieved using a Fabry-P\'erot cavity with $0.5\,$mm mirror spacing~\cite{Ritter2012}. 

Multiple cavity geometries have been utilized with neutral atoms (see Fig.~\ref{fig:cavities}). The size of the cavity can vary from centimeter scale all the way down to a few micrometer. One important figure of merit to characterize the coupling of a single atom to the cavity field is the cooperativity $C=g^2/\kappa\gamma$, where $g$ is the coupling strength between the atom and the cavity mode, $\kappa$ is the decay rate of the cavity and $\gamma$ is the decay rate of the atom~\cite{Reiserer2015}. For $C>1$ the emission of the atom into the cavity mode is strongly enhanced (Purcell enhancement), which is desired for increasing REG rates. All the cavity architectures shown in Fig.~\ref{fig:cavities} have been able to achieve large Purcell enhancements and in principle all of them could be used in a quantum network setting with QPUs consisting of many individually trapped atoms. For larger cavities with millimeter to centimeter sizes, arrays of atoms could be directly trapped inside the cavity (see also Fig.~\ref{Fig1} and~\ref{fig:processing-node})~\cite{Deist2022b,Periwal2021} while for smaller, micrometer-sized geometries the coherent motion of atoms in and out of the cavity mode could enable the integration with large processing arrays~\cite{Dordevic2021,Bluvstein2022} (see also section~\ref{sec:integration}). 

\subsection{Interfacing with telecom wavelengths}

A major bottleneck for long-distance quantum networking is the loss of single photons in optical fibers. This loss requires operation in the telecommunication wavelength band ($\approx1.25-1.65$ $\mu$m) where the attenuation length is maximal. At the optimal wavelength of 1550~nm the attenuation rate in state-of-the-art optical fibers is $\approx0.2$~dB/km - corresponding to an attenuation length
of about $\approx22$ km or an attenuation by a factor of 10 for a length of 50 km~\cite{Corning2020}. However, all atomic transitions that are currently used for quantum networking are at wavelengths much smaller than the ideal telecommunication case and thus suffer from significantly shorter attenuation lengths (e.g. at 780~nm as used in~\cite{Hofmann2012,Ritter2012} typical fiber attenuation is $\approx 4\,$dB/km corresponding to an attenuation length $\approx4\,$km). Therefore, for long-distance fiber links it is required to convert the emitted light to telecom wavelengths. 

One possible approach for telecom operation is to convert the photons that are emitted at shorter wavelengths to the telecom band. Nonlinear optics offers standard tools for wavelength conversion like sum-frequency- or difference-frequency-conversion, where the wavelength of the emitted photon is converted inside a nonlinear optical crystal with a strong pump laser to a telecom wavelength~\cite{NonlinOpt1991}. Compared to standard conversion applications, now one has to make sure that both basis states of the photonic qubit are converted. For example, due to the birefringence of the $\chi^{(2)}$-nonlinear crystal, this process only works for one polarization, which necessitates to run two conversion processes in parallel and coherently~\cite{Bock2018}. Particularly useful in this respect are Sagnac-type set-ups where, with a polarizing beam splitter at the input, the two polarizations propagate in opposite directions \cite{Ikuta2018}. Periodically poled crystals enable an overall conversion efficiency of more than $50\,\%$~\cite{Leent2020}. Narrowband filtering to reduce Raman background enabled the observation of entanglement between an atomic state and the polarization of a photon after 20~km of fiber~\cite{Leent2020} and to create REG between two atoms separated by 33 km of optical fiber~\cite{Leent2022}.

In section \ref{sec:direct-telecom} we will describe alternative approaches that enable direct telecom operation of the atom.

\subsection{Towards link efficiency \texorpdfstring{$>$}{TEXT} 1}
\label{sec:link-eff}

The functionality of a quantum network improves as entanglement between nodes becomes available a larger fraction of the time. The figure of merit which indicates exactly this availability is the link efficiency $\eta=\gamma_\text{ent}/\gamma_\text{dec}$, where $\gamma_\text{ent}$ is the entanglement rate between two nodes and $\gamma_\text{dec}$ is the decoherence rate of the entangled state~\cite{Hucul2015,Humphreys2018}. A link efficiency $\eta$ larger than one means that entanglement can be established faster than the shared quantum state decoheres. This is an outstanding challenge, especially for long-distance networks. $\eta$ larger than unity has already been demonstrated with trapped ions~\cite{Hucul2015} and NV centers in diamond~\cite{Humphreys2018}; however, only over short, meter-scale distances.

The REG protocols described above are designed to account for high photon loss due to finite collection efficiency and long fiber links. Yet,  inevitably the REG success rate  will be low for a low success probability per attempt combined with a limited attempt rate. 

For long fiber links, the latter is set by the two-way communication time $t_\text{comm}$ necessary to send the entangled photons to the BSM station and the classical heralding signal back to the nodes.
For instance, for a fiber length of $L$ km between repeater stations and with the Bell-state measurement half-way in between, the two-way communication time is $t_\text{comm}=L\times 5 \mu s$ (given that the speed of light in the fiber is about $2 \times 10^8\,m/s$). The maximum rate at which the excitation events can be repeated is then $\gamma=1/t_\text{comm}$, i.e. for a fiber length of $L = 20\,$km we obtain a minimum time between the attempts of $t_\text{comm}=100\,\mu $s and a maximum attempt rate of only $\gamma=10\,$kHz.
In addition, for large networks with many intermediate repeater stations and success probability less than one, not all links can be connected simultaneously. Therefore, some shared entanglement must be coherently stored while other REG attempts are ongoing, thus requiring even lower decoherence rates $\gamma_\text{dec}$. 

To maximize the success probability per attempt, we must maximize the photon collection efficiency. It thus will be crucial to apply measures like, as mentioned above, placing the atom in an optical cavity~\cite{Reiserer2022}, \textit{and} maximizing the attenuation length in optical fiber by operating at the optimal wavelength (see also section~\ref{sec:nextsteps}). In addition, entanglement rates between distant nodes have to be greatly enhanced by using multiple atoms per node (see next section). Instead of waiting for the two-way communication for each individual attempt, multiplexing REG with a burst of attempts during $t_\text{comm}$ will enable $\eta >1$.

\section{Multiqubit nodes and processing}\label{sec:Processing} 
In parallel to the progress using individual atoms as network nodes, there have been major advances in using individual atoms as building blocks for quantum processors and simulators~\cite{Morgado2021,Browaeys2020}. The enabling technology for these developments are optical tweezer arrays that are being used to assemble atomic qubit arrays with hundreds of single atoms in arbitrary geometries~\cite{Scholl2021,Ebadi2021}. 
While all demonstrations of REG with neutral atoms to date have used only a single atom at each node, the realization of the envisioned networking operations highlighted in Section~\ref{vision} will require many qubits per node. This includes, for example, protocols that utilize local deterministic operations for entanglement swapping at QRs, or larger scale QPUs that are linked together for distributed quantum computation. A first step towards scaling the number of atoms per node is ref.~\cite{Langenfeld2021b} which used two atoms in a single node to distribute entangled photons.  

In this section we will give a brief review of the progress on atom array QPUs and discuss the prospect of combining such arrays with optical interfaces for REG. Table~\ref{tab:processing} gives an overview of key metrics that have been achieved in the context of neutral atoms for quantum information processing and quantum network demonstrations.

\begin{table}
\begin{tabular}{|l|l|l|}
\hline

\parbox[t]{7mm}{\multirow{11}{*}{\rotatebox[origin=c]{90}{\makecell{Processing \\ capabilities}}}} & Qubit number & $>$200~\cite{Ebadi2021,Scholl2021} \\\cline{2-3}
 & Lifetime $T_1$ & $>$minutes \\\cline{2-3}
 & Coherence $T_2$ & $40\,$s~\cite{Barnes2022} \\\cline{2-3}
 & Gate time & $<\mu$s for 1Q and 2Q\\\cline{2-3}
 & Readout speed & \makecell[l]{$\sim$1-10$\,$ms (free space), \\ $\sim$10-100$\,\mu$s (cavity)} \\\cline{2-3}
 & \makecell[l]{Mid-circuit \\readout} & \makecell[l]{dual-species~\cite{Singh2022},\\ cavity-based~\cite{Deist2022c}} \\\cline{2-3}
 & Single-qubit gate & \makecell[l]{$\sim$99.96\% (global), \\$>$99.6\% (selective)~\cite{Xia2015,Wang2016,Ma2022}} \\\cline{2-3}
 & Two-qubit gate & 96-99\%~\cite{Levine2019,Madjarov2020}\\\cline{2-3}
 & Three-qubit gate & $>$87\%~\cite{Levine2019}\\ \hline
\parbox[t]{10mm}{\multirow{12}{*}{\rotatebox[origin=c]{90}{\makecell{Network \\ capabilities}}}} & \makecell[l]{Atom-atom \\REG $\mathcal{F}$} & \makecell[l]{89\%~\cite{Zhang2022}, \\
98.7\% (postselected)~\cite{Ritter2012}}\\\cline{2-3}
 & \makecell[l]{Atom-atom \\REG rate} & \makecell[l]{30~mHz (400~m)~\cite{Rosenfeld2017},\\
 30~Hz (21~m)~\cite{Ritter2012}}  \\\cline{2-3}
 & Telecom operation & \makecell[l]{conversion~\cite{Leent2022},\\ direct${}^*$~\cite{Uphoff2016,Covey2019b,Menon2020}}  \\\cline{2-3}
 & Cavity integration & \makecell[l]{nanophotonic: $C\sim70$~\cite{Samutpraphoot2020},\\ fiber: $C>200$~\cite{Barontini2015},\\ Fabry-P\'erot ($\sim$mm): $C=7.7$~\cite{Daiss2021},\\ near-concentric: $C\sim5$~\cite{Davis2019,Deist2022b}} \\ \cline{2-3}
 & Ion-Ion REG $\mathcal{F}$ & 96\%~\cite{Nadlinger2022} \\\cline{2-3}
 & Ion-Ion REG rate & \makecell[l]{$182\,$Hz ($2\,$m)~\cite{Stephenson2020}, \\$0.43\,$Hz ($230\,$m)~\cite{Krutyanskiy2023}} \\ \hline
\end{tabular}
\caption{\label{tab:processing} Key characteristics and figures of merits for processing with arrays of atomic qubits and quantum networking. For comparison REG rate and fidelity for trapped ions is shown. Asterisks indicate proposals. For reference, ion-ion entanglement rates and fidelities are also included.}
\end{table}

\subsection{Arrays of atomic qubits}
A tightly focused laser beam, called optical tweezer, with a wavelength that is far off resonant and red detuned from an optical transition of the atom creates a trapping potential~\cite{Grimm2000}. At foci of $\sim 1\,\mu$m only a single atom will be trapped as light-assisted collisions lead to two-body losses resulting in a probabilistic loading process with either one or zero atoms being captured in the tweezer~\cite{Schlosser2001}. Large arrays of hundreds of optical tweezers can be created by using light-shaping or beam-steering techniques with spatial light modulators (SLMs), digital mirror devices (DMDs) or acousto-optic deflectors (AODs). However, due to the probabilistic loading fractions of the atom arrays are typically on the order of $50-60\,$\%. In 2016, three research groups have demonstrated a rearrangement protocol that, by moving randomly loaded rubidium atoms, results in defect-free atomic arrays~\cite{Endres2016,Barredo2016,Kim2016}. Subsequently, such atom arrays have been created with hundreds of atoms in arbitrary 2D geometries~\cite{Scholl2021,Ebadi2021} and at lesser numbers in 3D geometries~\cite{Barredo2018}. Furthermore, the optical tweezer technique has been extended to alkaline-earth atoms, such as strontium and ytterbium~\cite{Cooper2018,Norcia2018b,Saskin2019}, and individual molecules~\cite{Anderegg2019}. These arrays form the basis of quantum simulators~\cite{Browaeys2020}, quantum processors~\cite{Morgado2021}, optical atomic clocks~\cite{Madjarov2019,Norcia2019,Young2020}, and we envision them as the QPUs in a quantum network architecture.

\subsection{Gate operations and coherence}
Long-lived qubit states in atomic arrays can be conveniently defined using spin states, such as hyperfine states in alkali atoms~\cite{Xia2015,Wang2016} or nuclear spin states in alkaline-earth atoms~\cite{Ma2022,Barnes2022,Jenkins2022}. The coherence times of these states are on the order of several to tens of seconds~\cite{Wang2015,Barnes2022}, exceeding typical gate operation times ($\sim\mu$s) by more than six orders of magnitude. Single qubit operations can be implemented by either using microwave or Raman manipulations and gate fidelities have reached up to 99.96~\% for global control of all the atoms within the array~\cite{Ma2022} and $>$99.6\% for site selective control of single atoms within the array~\cite{Xia2015,Wang2016}.

Photon-mediated entanglement protocols between two atoms in an optical cavity have been realized~\cite{Pellizzari1995,Sorensen2003,Welte2017,Welte2018,Dordevic2021}, but are hampered by spontaneous emission and cavity loss. Therefore, two-qubit gates are typically realized by relying on direct interactions between the atoms. At separations of a few micrometers, interactions between the atoms in their ground states are completely negligible. However, coupling atoms to highly excited states - Rydberg states - leads to strong interactions that can be used for high-fidelity 2-qubit gate operations~\cite{Jaksch2000,Saffman2010}. The fidelities of these operations are above 97\%~\cite{Levine2019,Graham2019} with a demonstration of an entangling operation between two strontium atoms of larger than 99\%~\cite{Madjarov2020}.

Single-qubit and two-qubit gates are now being used to execute quantum algorithms on atomic arrays~\cite{Bluvstein2022,Graham2022}. The ability to integrate such processing arrays with photonic links would vastly expand the opportunities for quantum networks. 

\subsection{Qubit measurements}

Atomic qubit states are typically measured using state-dependent losses in combination with fluorescence detection of the atoms. Here, a strong resonant `push beam', that is resonant with a cycling transition of one qubit state but not the other, expels atoms in the resonant state, while the atoms in the off-resonant state remain trapped. A subsequent fluorescence image of the atoms reveals which atoms have been lost and therefore the qubit states prior to the push-beam can be deduced. While this detection technique can have very high fidelity it is highly destructive and basically marks the end of the protocol. 

Lossless state detection has been implemented by performing state-dependent fluorescence. Crucially, this requires high photon collection efficiency in order to collect enough signal before the atoms are lost or qubit states are accidentally changed. Using high-numerical-aperture lenses or microscope objectives a detection fidelity of $>$99\% has been achieved with a probability of atom survival of $>$98\%~\cite{Dorantes2017,Kwon2017,Covey2019}.

Still, this measurement technique is insufficient for many quantum information and networking protocols as all the atoms in the array are measured at the same time and the measurement time is typically larger than the coherence time. Looking back to figure~\ref{Fig1}C it is apparent that the ideal measurement needs to be atom-selective and at the same time it should not decohere the atoms that are not being measured. Such ``mid-circuit measurements" are challenging since fluorescence beams typically excite the whole array and cause decoherence of all the atoms. 

There are several approaches that have recently been introduced to measure selected atoms without affecting the coherence of the remaining ones. The first technique uses a readout cavity in combination with coherent motion of the atoms. The atom that is to be measured, is moved into the cavity which provides a readout zone. Cavity-enhanced fluorescence of the atom in this zone reveals the qubit state without affecting the atoms outside the cavity~\cite{Deist2022c}. Alternatively, the resonance frequency of the cavity can depend on the state of the atom such that light reflecting off the cavity yields information about the atomic state~\cite{Gehr2010,Bochmann2010,Volz2011}. 

A second technique uses dual-species atomic arrays consisting, for instance, of individual rubidium and individual cesium atoms. Here the large frequency difference between the two atomic species, enables the readout of one species without affecting the coherence of the other species~\cite{Singh2022}. The readout of the auxiliary qubits can then be used for measurement-based protocols, such as error correction, or entanglement distillation. Finally, a third technique uses shelving states in which atoms can be selectively ``hidden" from the readout light~\cite{Chen2022,Wu2022}. This technique is already routinely being used for trapped ions~\cite{Monz2016,Erhard2021}, and is perhaps best suited for alkaline earth atoms that offer long-lived optically excited states although it has recently been demonstrated with cesium~\cite{Graham2023}.

\subsection{Integrating atomic arrays with optical interfaces}
\label{sec:integration}

\begin{figure*}[ht!]
    \centering
    \includegraphics[width=13cm]{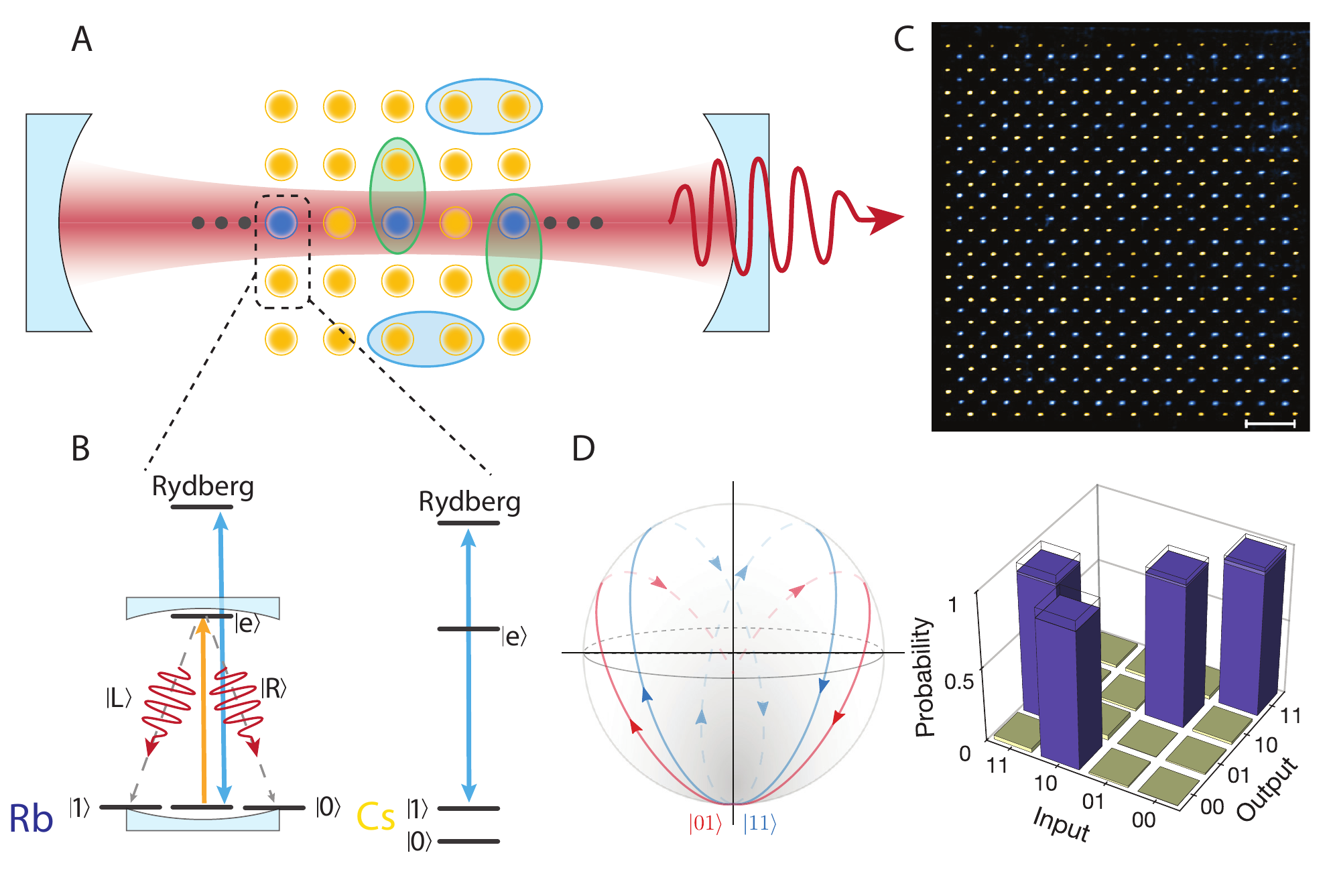}
    \caption{
        \textbf{Integrating photonic links with processing arrays.} \textbf{A} An atomic array is placed in a macroscopic, near-concentric optical cavity. Blue circles indicate communication qubits with optical resonances that couple to the cavity. Yellow circles indicate data qubits of a different atomic species that do not couple to the cavity. Green ovals represent inter-species two-qubit gates enabled by Rydberg interactions used to transfer entanglement from the communication to the data qubits. Blue ovals represent intra-species two-qubit gates on the data qubits. \textbf{B} Relevant level structure. Shown are long-lived hyperfine states as qubit states, first excited states $\ket{e}$ for REG and readout, as well as highly excited Rydberg states for interactions. Using a dual-species architecture, for instance consisting of rubidium (Rb) and cesium (Cs), allows for independent control of these states due to the large frequency separation between the two species. \textbf{C} Average fluorescence image of a dual-species atom array. Scale bar indicates $20\,\mu$m. Adapted from~\cite{Singh2021}. \textbf{D} A two-qubit gate via Rydberg interactions. In the gate protocol of Levine et al.~\cite{Levine2019} choosing certain detunings and phases of the Rydberg excitation leads to different closed loop evolution of the input states on the Bloch sphere. This can be used to implement a controlled-phase gate that in combination with single-qubit manipulations creates a Bell state. Adapted from~\cite{Levine2019,Ma2022}.
    }
    \label{fig:processing-node}
\end{figure*}

There are multiple promising strategies for combining atomic processing arrays with photonic interfaces. In figure~\ref{fig:processing-node}~A we consider an architecture using a near-concentric, centimeter-size optical cavity in which an array of atoms can be trapped at the center of the cavity. 
Cooperativities of $C > 1$ have been demonstrated in such cavities~\cite{Davis2019,Deist2022b}. To ensure that the cavity is not constantly coupling to all the atoms, as mentioned above, selected atoms can either be shelved into states, that are off-resonant with the cavity mode~\cite{Chen2022,Wu2022} or a dual-species architecture can be used, in which the optical frequencies of the two types of atoms are distinct and the cavity is only on resonance with one species~\cite{Young2022b} (Fig.~\ref{fig:processing-node}). One species of atoms then acts as the communication qubits, which are used to distribute entanglement between distant nodes, while the other species acts as the data qubits that can further store and process quantum information. Figure~\ref{fig:processing-node}~B shows a fluorescence image of an array of individual cesium and rubidium atoms with 512 trapping sites which can be independently controlled and measured due to the large frequency difference~\cite{Singh2021,Singh2022}. Two qubit operations between the communication and data qubits can be achieved using Rydberg gates (Fig.~\ref{fig:processing-node}~D) by which the entanglement between distant nodes can be transferred onto the data qubits and used in further processing. 

An alternative integration strategy would be the movement of communication qubits in and out of the cavity coupling region. Remarkably, the coherence of the atoms can be maintained during such movement and recent demonstrations have shown the compatibility with nanophotonic crystal cavities~\cite{Dordevic2021} and Fabry-P\'erot cavities~\cite{Deist2022b}. The processing array would then be placed at a distance from the cavity, which has the added benefit of avoiding detrimental effects on the Rydberg gates from stray electric fields originating from surface charges on the cavity structures. A recent demonstration of an entangling operation between two atoms placed $\sim100\,\mu$m away from a nanophotonic crystal cavity shows the viability of this approach~\cite{Ocola2022}

\section{Next steps}\label{sec:nextsteps}
We believe that the stage is set for several enabling advances in quantum networking with arrays of individual neutral atoms, which will improve network REG rates as well as network complexity. In the following we highlight these enabling advances as next steps that are currently within experimental reach.

\subsection{Direct telecom operations}
\label{sec:direct-telecom}
There is growing interest in performing REG directly in the telecommunication wavelength band with suitable atomic species and transitions~\cite{Uphoff2016,Covey2019b,Menon2020}. We highlight two promising approaches.

\begin{figure*}[t!]
    \centering
    \includegraphics[width= 14cm]{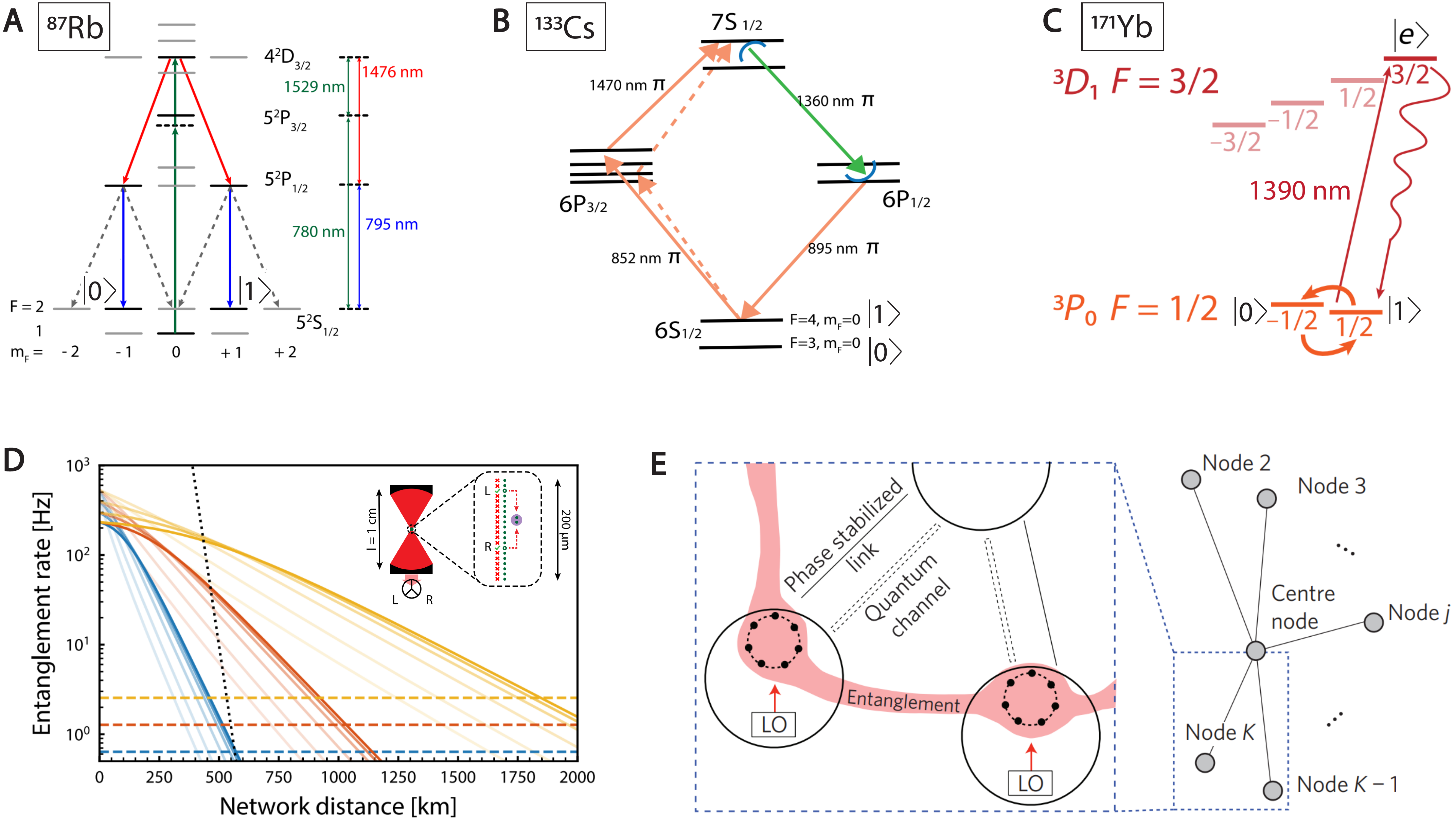}
    \caption{
        \textbf{New opportunities for quantum networking with neutral atom arrays}. \textbf{A} Telecom-band polarization encoding using a multi-level protocol in Rb. Adapted from~\cite{Uphoff2016}. \textbf{B} Telecom-band time-bin encoding using a multi-level protocol in Cs. Adapted from~\cite{Menon2020}. \textbf{C} Telecom-band time-bin encoding using a two-level protocol from a long-lived metastable state in $^{171}$Yb. Adapted from~\cite{Covey2019b}. \textbf{D} Entanglement rates in a network with three (blue), seven (orange), and 15 (yellow) intermediate repeater stations. The opacity scale indicates the number of atoms ($N=200$ is fully opaque) used in a multiplexing protocol. The inset shows a vision for realizing multiplexing protocols. Adapted from~\cite{Huie2021}. \textbf{E} A quantum network of atomic clocks. Entanglement is distributed within and between nodes. Phase stabilization of the local oscillators is also required. Adapted from~\cite{Komar2014}.
    }
    \label{Fig5}
\end{figure*}

Strong telecom-band transitions between short-lived excited states can be used in a multi-photon process that mitigates unwanted emission during the REG protocol. Two examples are shown in Fig.~\ref{Fig5}A and~\ref{Fig5}B for rubidium and cesium, respectively. Figure~\ref{Fig5}A shows a polarization encoding approach from Ref.~\cite{Uphoff2016} where the telecom photon polarization (red transition, with a wavelength of 1475 nm and an attenuation of 0.24 dB/km) is entangled with the Zeeman state of the $F=2$ hyperfine ground state. The other emitted photon (blue transition) has the same polarization in each case, and can be collected with a second optical cavity mode (see Fig.~\ref{fig:cavities}B) and detected as part of the heralding process for remote entanglement generation. Figure~\ref{Fig5}B shows a time-bin encoding approach from Ref.~\cite{Menon2020} with similar level structure. The four-level diamond process uses fine-tuned temporal pulse control to result in emission into the cavity mode with minimal loss from the short-lived intermediate states. The process leaves the atom in the same state in which it started, as described above. Similar schemes are widely used for microwave-optical conversion with atomic and atom-like systems~\cite{Lauk2020,Kumar2023}.

Instead of relying on these relatively advanced schemes to employ transitions between short-lived excited states, it may be possible in rare cases to identify suitable transitions from a long-lived metastable excited state. In particular, neutral ytterbium (Yb) -- which is alkaline-earth-like, meaning it has two valence electrons -- has a strong transition ($\Gamma/2\pi=320$ kHz) at 1389 nm (with $\sim$0.35 dB/km attenuation) from its long-lived metastable ``clock" state. The clock state has a lifetime of $\approx20$ seconds, and the nuclear spin degree of freedom ($I=1/2$ for $^{171}$Yb) can be used as a qubit. Figure~\ref{Fig5}C shows a time-bin encoding approach from Ref.~\cite{Covey2019b} that operates essentially like a two-level system. This approach requires initialization in the ``clock" state for each attempt which introduces some complexity overhead, but this can be done quickly ($\gtrsim200$ kHz) via the anticipated ability to directly drive the ``clock transition" for tweezer-trapped atoms~\cite{Jenkins2022,Chen2022} or by incoherent pumping~\cite{Madjarov2020}. Nuclear spin rotations can be performed with fast Raman pulses ($\Omega/2\pi\approx1$ MHz)~\cite{Jenkins2022}. There is currently intense interest in using this metastable nuclear spin qubit for quantum computing and metrology~\cite{Jenkins2022,Ma2022,Chen2022,Wu2022}, and the extension to networking is very natural.

\subsection{Multiplexed networking with atom array nodes}
Atom arrays offer opportunities for networking beyond merely the access to a large number of data qubits. An array of communication qubits at each node and individual control therein can be used to improve the REG rates over the links and to create larger numbers of entangled pairs per link. We now discuss both opportunities. Note that the optimal scheme may be different between Fabry-P\'erot and nanophotonic cavities.

\subsubsection{Enhanced REG rates}
Long-distance quantum networking is plagued by low success probabilities and high latency for heralded success. As described in section~\ref{sec:link-eff}, REG rates for long links are limited almost entirely by communication timescales -- set by the speed of light. Therefore, a crucial goal is to increase the success per attempt, which can be accomplished by attempting to create entanglement in parallel with many atoms at each node. Many variants of this multiplexing theme have been proposed that are based on myriad degrees of freedom: polarization~\cite{Graham2013}, frequency~\cite{Sinclair2014,Wengerowsky2018}, and time~\cite{Kaneda2015}. Atom arrays are naturally suited for temporal multiplexing, where the atom-photon entanglement protocol can be performed selectively on each atom in succession in a coordinated fashion between the two nodes. Such a protocol has been considered in detail in Ref.~\cite{Huie2021}, and it was shown that the time required to multiplex over an array of $N\approx200$ atoms can still be shorter than $t_\text{comm}$, the latency of communication over a long link ($L\gtrsim40$ km). In this regime, the success probability per attempt essentially scales linearly with $N$. For much larger $N$, the time required for REG operations on each atom becomes prohibitively expensive. The heralding process records information of which atom(s) successfully created a Bell pair, and then they can be re-positioned as needed for subsequent operations (see inset to Fig.~\ref{Fig5}D). Figure~\ref{Fig5}D shows expected results with many intermediate QRs and demonstrates that link efficiencies above one (indicated by the horizontal dashed lines) can be achieved even for distances on the order of 1000 km~\cite{Huie2021}. 

\subsubsection{Distributing multiple entangled pairs}
The above multiplexing protocol also enables the generation of more than one Bell pair across the links. As discussed in Section~\ref{vision}, at least two Bell pairs across each link are needed for entanglement purification~\cite{Bennett1996a,Dur2003,Kalb2017} (see Fig.~\ref{Fig1}C), which will be necessary due to the decrease in fidelity associated with connecting progressively more links. Moreover, multiple Bell pairs can also be used for more advanced functions such as fault-tolerant networking and distributed computing. Multiplexing can yield a large number of Bell pairs spanning metropolitan-scale links with moderately high rates~\cite{Huie2021}. With five or more Bell pairs, fault-tolerant protocols such as surface codes~\cite{Fowler2012} can be realized with stabilizers spanning the network links~\cite{Nickerson2013,Nickerson2014,Auger2017}.  

\section{Outlook on future applications} \label{sec:outlook}
Advanced, long-distance networks offer exciting new opportunities for information processing and beyond. As an outlook, we will highlight a few such opportunities -- many of which leverage the unique properties of neutral atom arrays.

\subsection{Efficient quantum communication}
Quantum repeater protocols enable sharing of entanglement with only polynomially increasing resources depending on the distance between the users. This is in stark contrast to current state-of-the-art fiber-based quantum key distribution (QKD) where the achievable key rate necessarily decreases exponentially with the distance. Thus, these systems must rely on so-called trusted nodes, where links with a typical distance of 70-100 km are bridged with high key rates and where the classical keys at these nodes are combined to create a new key between the end nodes~\cite{Pirandola2019}. Yet, an attack on a single trusted node suffices to learn the final key along a link. Instead, shared entanglement across a future quantum network provides maximal secrecy over any distance. Moreover, as the state of atoms at the QPUs can be measured with effectively unit detection efficiency, fully device-independent QKD can be performed giving the ultimate security not only against attacks on the quantum channel, but also against misalignment and manipulations of the devices by the eavesdropper~\cite{Zapatero2023}.

In quantum networks, entanglement can not only be distributed between two parties but also between several parties, thereby allowing one to share any multi-party entangled state across the partners of the network. This enables, e.g., the creation of secure keys between several parties~\cite{Tittel2001,Chen2005,Schmid2005} which in turn form the basis for secret sharing or conference agreement between several parties~\cite{Murta2020}. Among others, novel schemes with higher efficiency or schemes which are not even possible with only classical communication become feasible, i.e., a scheduling (or Byzantine) agreement~\cite{Fitzi2001,Gaertner2008,Taherkhani2017}.

\subsection{Distributed and blind quantum computing}
Multi-node networks with QPUs at each node and many Bell pairs shared on each link enable distributed quantum computing protocols. It is believed that modular architectures~\cite{Monroe2014} for quantum computing may enhance the prospects for scaling QPUs to the level required for fault-tolerant operations~\cite{Knill1998,Knill2005,Fowler2012} capable of addressing questions of societal relevance. Moderate-scale QPUs with $N\approx100-1000$ atoms at each node can be ``wired" together to reach a ``quantum supercomputer" that distributes the computation among the nodes. This architecture also offers the ability to perform ``blind" computing ~\cite{Broadbent2009,Fitzsimons2017} operations in which the user can access a remote quantum computer without relying on classical information transfer between the user and the computer, as is the case in cloud-based quantum computing schemes. The no-cloning theorem~\cite{Wootters1982,Buzek1996} guarantees that eavesdropping and hacking is not possible if the remote quantum computer is accessed with only quantum channels such as those offered by a quantum network.

Crucially, it has been shown that the fidelity requirements for the links between nodes are not as stringent as the requirements for operations within the nodes~\cite{Jiang2007,Nickerson2013,Nickerson2014,Ramette2023}. Table~\ref{tab:processing} shows that state-of-the-art experiments are well matched with these requirements: the record REG fidelities currently reach $\approx0.90$ while the deterministic local entanglement fidelities are $\approx0.99$. It has been shown that fault-tolerant operation may succeed in staying below the associated error threshold with $\sim1$\% errors within the nodes (bulk) and $\sim10$\% errors between nodes (interface)~\cite{Nickerson2013,Nickerson2014,Ramette2023}. These results suggest that fault-tolerant scaling of error-corrected modular devices is within reach when combining existing capabilities.

\subsection{Networked clocks and sensors}
Opportunities for advanced quantum networks that go beyond information processing include single-photon astronomy~\cite{Rauch2018}, studies of quantum foundations~\cite{Hensen2015}, sensing~\cite{Malia2022b}, and even timekeeping~\cite{Komar2014}. Optical atomic clocks based on ensembles of alkaline-earth(-like) atoms have reached record precision~\cite{Ludlow2015} that now approaches $10^{-19}$ at one second of averaging. Quantum networks of such clocks~\cite{Komar2014,Komar2016,Nichol2022} could offer a distributed time standard with precision enhanced by quantum entanglement~\cite{Gil2014,Kessler2014,Pezze2018} and secured by the no-cloning theorem~\cite{Wootters1982,Buzek1996}. Recently, a two-node quantum network of single-ion clocks has been realized~\cite{Nichol2022}, and we anticipate an exciting future of clock networks involving many nodes of atom array optical clocks~\cite{Madjarov2019,Norcia2019,Young2020,schine2022}. Beyond time keeping, optical atomic clocks have become exquisite quantum sensors, and are now able to detect gravitational redshifts at the millimeter distance scale~\cite{Bothwell2022,Zheng2022,Zheng2022b}. Our envisioned network offers exciting opportunities for distributed sensing that could contribute to the search for dark matter~\cite{Derevianko2014,Wcislo2017,Wcislo2018}, gravitational waves~\cite{Kolkowitz2016}, and other fundamental physics.

\section{Summary}\label{sec:summary}
In this Perspective, we have reviewed the rapid progress on quantum networking with individual neutral atoms based on efficient light-matter interfaces, and quantum computing via Rydberg-mediated interactions in arrays of neutral atoms. Currently, novel systems are being designed and built in an effort to combine several recent developments into a single apparatus. The confluence of these research areas enhances the feasibility of developing long-distance networks with nodes that comprise advanced quantum processors. We look forward to the exciting progress toward this vision over the coming years and decades. We believe that neutral atom processing nodes will play a large role in this future, possibly supported by disparate hardware platforms to leverage their unique strengths.

\section*{Acknowledgments}
We thank Andreas Reiserer and Johannes Borregaard for critical reading of the manuscript. JPC and HB acknowledge funding from the NSF QLCI for Hybrid Quantum Architectures and Networks (NSF award 2016136) and the U.S. Department of Energy, Office of Science, National Quantum Information Science Research Centers. JPC acknowledges funding from the NSF PHY Division (NSF award 2112663), the NSF Quantum Interconnects Challenge for Transformational Advances in Quantum Systems (NSF award 2137642), the ONR Young Investigator Program (ONR award N00014-22-1-2311). HB acknowledges funding from the Office for Naval Research (Grant No. N00014-20-1-2510) and the NSF Quantum Interconnects Challenge for Transformational Advances in Quantum Systems (NSF award 2138068). HW acknowledges funding from the German Federal Ministry of Education and Research [Bundesministerium f\"ur Bildung und Forschung (BMBF)] within the project QR.X (16KISQ002).
\vspace{5mm}

\bibliographystyle{h-physrev}
\bibliography{library,HB_refs,HB_refs2}

\end{document}